\documentclass[final,1p,times]{elsarticle} 
\usepackage{graphicx} 
\usepackage{amssymb} 
\usepackage{amsthm} 
\usepackage{lineno} 

\journal{Nuclear Physics A} 
\begin{document} 

\begin{frontmatter} 


\title{Modification of jet-like correlations in Pb-Au at the SPS}

\author{Harald Appelsh\"auser$^{a}$ and Stefan Kniege$^{a}$  for the CERES Collaboration}

\address[a]{Institut f\"ur Kernphysik, Johann Wolfgang Goethe-Universit\"at Frankfurt, 60438 Frankfurt am Main, Germany}

\begin{abstract} 
A high statistics study of high-$p_t$ two-particle azimuthal 
correlations in Pb-Au at 
$\sqrt{s_{\rm NN}}=17.2$~GeV, performed by the CERES experiment at 
the CERN-SPS, is presented. 
A broad away-side correlation with significant dip at 
$\Delta\phi\approx\pi$
is observed. The shape and magnitude of the correlation is similar to 
measurements at RHIC. In comparison to PYTHIA calculations, we 
observe a significant excess of soft particles at the away-side.
A study of charge correlations between trigger and associated particles
disfavors vacuum fragmentation of the away-side jet and suggests 
significant energy transfer of the hard-scattered parton to the
medium.

\end{abstract} 

\end{frontmatter} 

\linenumbers 

The modification of jet properties in heavy-ion collisions at RHIC 
has been interpreted by a strong final state interaction of 
hard-scattered partons with the surrounding matter. 
In particular, it has been argued that the strong suppression of particles
at high transverse momentum $p_t$ is consistent with energy loss of
a parton in a QGP.
Such modifications show a distinct dependence on the 
$p_t$ range under study. At moderate $p_t$, the away-side jet correlation 
exhibits significant broadening and possibly a two-peak structure which
may be a manifestation of the dissipated parton energy in the medium. 
The so-called {\em medium response} is of particular interest because
it may reflect properties of the medium created in heavy-ion collisions,
such as the velocity of sound.
In this context it is of utmost importance to understand the collision
energy dependence of such modifications as they should be sensitive
to qualitative changes of the medium properties, such as the creation
of a QGP~\cite{review}.

At the SPS, pioneering studies
on jet modifications in Pb-Au collisions have been reported by CERES,
albeit with considerable statistical uncertainties~\cite{ceres1}. 
In this presentation,
we report on results of a recent analysis of a high statistics Pb-Au
data sample at 158$A$~GeV/$c$~\cite{ceres2,kniege},
recorded with the CERES Time Projection Chamber~\cite{ceres3}. 

We study jet-like correlations by measuring, for a trigger particle
and all associated particles in an event, the distribution $S(\Delta \phi)$
where $\Delta \phi=\phi_1-\phi_2$ is the difference in the azimuthal angle
between trigger and associated particle. Within a given event, the trigger
is the particle of highest $p_t$ within a defined trigger $p_t$ range. 
Non-uniformities in the detector acceptance are accounted for by division
by a mixed-event distribution $B(\Delta \phi)$, where trigger and associate
particle are taken from different events. The normalized correlation function
\begin{equation}
C_2=\frac{\int B(\Delta\phi' d(\Delta \phi')}{\int S(\Delta\phi' d(\Delta \phi')}
\cdot \frac{S(\Delta \phi)}{B(\Delta \phi)}
\end{equation}
contains correlations due to jets and elliptic flow:
\begin{equation}
        C_2(\Delta \phi) = J(\Delta \phi)+
                b\cdot(1+2\langle v_2^T \rangle 
                \langle v_2^A\rangle \cos(2\Delta \phi)), 
\end{equation}
where $\langle v_2^T \rangle$ and $\langle v_2^A\rangle$ are the average
elliptic flow coefficients determined in the trigger and associate
$p_t$ range, respectively. Assuming that the jet yield vanishes at its
minimum (the ZYAM conjecture~\cite{phe5,ZYAM}),
the elliptic flow contribution is adjusted by variation of $b$. 
We obtain the conditional yield $ \hat{J}_2(\Delta \phi) $
as the number of jet-associated particles
per trigger:
\begin{equation}
  \hat{J}_2(\Delta \phi) \equiv \frac{1}{N_T}
        \frac{dN_{TA}}{d\Delta\phi}=  \frac{1}{\epsilon} 
  \frac{J(\Delta\phi)}{\int{C_2(\Delta \phi')d(\Delta \phi')}}
  \frac{N_A}{N_T},
\end{equation}
where $N_T$ and $N_A$ are the total numbers of triggers and associates,
$N_{TA}$ is the number of jet-associated particles with the trigger
after subtraction of the flow-modulated background,
and $\epsilon$ the single track efficiency.
See~\cite{ceres2,kniege} for more details on the analysis. 

\begin{figure}[t]
\centering
\includegraphics[width=1.0\textwidth]{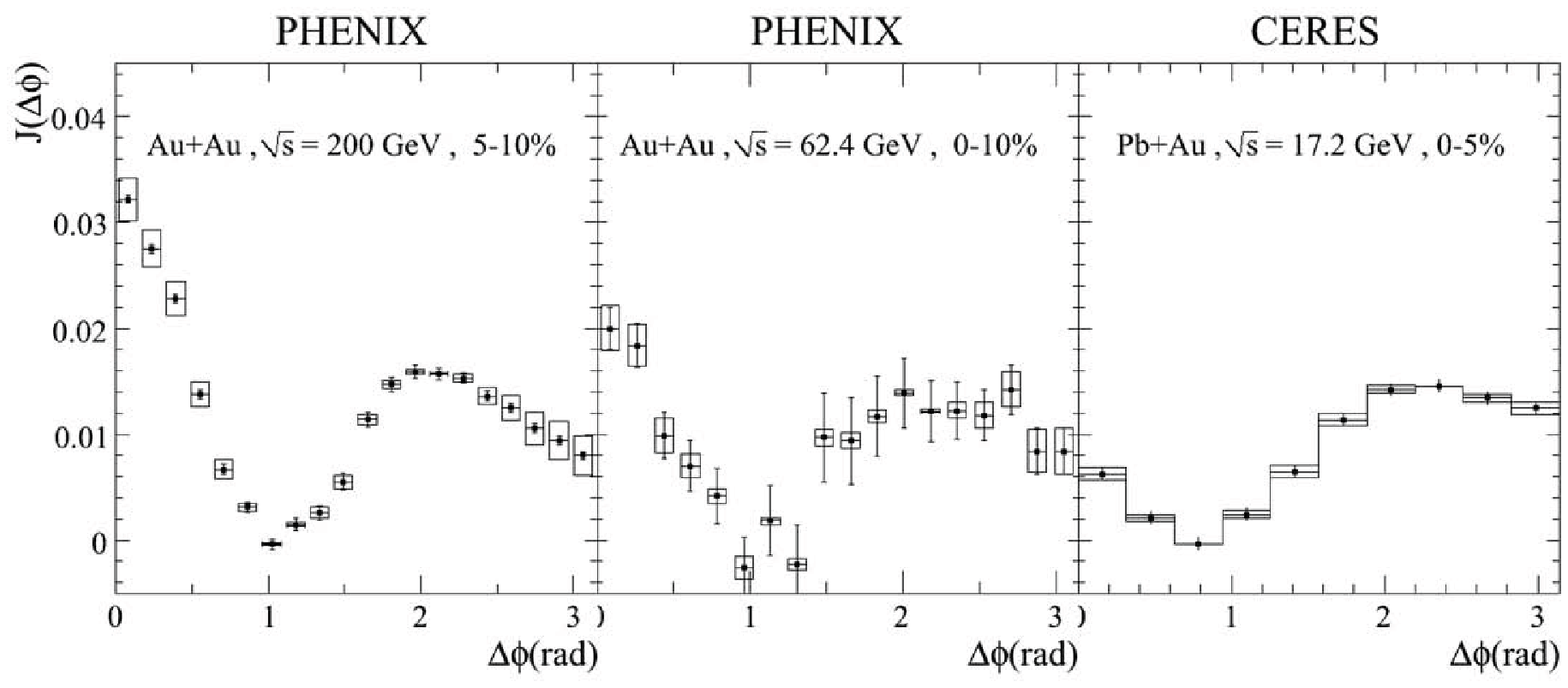}
\caption[]{$J(\Delta\phi)$ in central events
at different collision energies~\cite{ceres2,jia}. 
The trigger and associate $p_t$ ranges
are $2.5 < p_{t}(T) < 4.0$~GeV/$c$
and $1.0 < p_{t}(A) < 2.5$~GeV/$c$, respectively.}
\label{fig1}
\end{figure}

In Fig.~\ref{fig1} the correlation function after subtraction of the
flow contribution, i.e. the jet-like part $J(\Delta\phi)$, in central Pb-Au 
collisions is compared to results from Au-Au at RHIC~\cite{jia}. 
At the near-side,
the correlation is gradually decreasing as the c.m.s. energy is lowered. 
This can be explained
in terms of a steeper jet spectrum at the SPS, leading to less
associated particles for a given trigger $p_t$.
At the away-side, there is little variation of the correlation
strength with $\sqrt{s_{\rm NN}}$. At all
energies a broad correlation of similar magnitide with significant 
dip around 
$\Delta\phi\approx\pi$ is observed, which is in contrast to observations
in pp. The occurence of a seemingly universal response of the medium 
to a trigger particle of given $p_t$ in central nucleus-nucleus collisions
suggests that the properties
of the medium created at different collision energies exhibit surprising
similarities in terms of opaqueness to energetic partons.

\begin{figure}[t]
\centering
\includegraphics[width=1.0\textwidth]{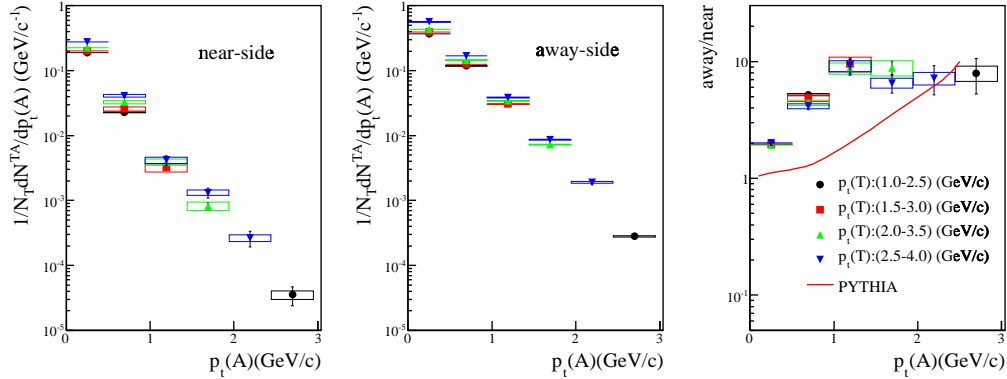}
\caption[]{Jet-associated yield as function of associated $p_t$ 
at the near-side (left) and away-side (middle). The ratio away/side
compared to PYTHIA is shown in the right panel (from~\cite{ceres2}).}
\label{fig2}
\end{figure}

In Fig.~\ref{fig2} the jet-associated yield in different ranges
of the trigger and associated $p_t$ is shown. 
Associated particles at $|\Delta\phi| < 1$ have been assigned to the 
near-side, and $1<|\Delta\phi|<2\pi-1$ to the away-side.
We observe that
the spectrum at the near-side is significantly softer than at the away-side.
This is a consequence of the trigger requirement of an energetic
leading particle at the near-side, leaving less energy for the associated
particles.
The away-side is not biased by the trigger condition, hence leading 
to a harder spectrum of the
associated particles.
The effect of the trigger bias is also observed in PYTHIA simulations,
shown as the ratio of yields away/near in Fig.~\ref{fig2} (right panel).
This qualitative agreement corroborates the fragmentation picture.
In the data, however, we observe a significant excess of soft particles
with $p_t < 2$~GeV/$c$ compared to the PYTHIA calculation, indicating 
significant medium modification of the fragmentation spectrum at the 
away-side. We have verified that the shape of the away-side spectrum
is undistinguishable from the inclusive $p_t$ distribution.

In the following, electric charge correlations between the trigger and
associated particles are investigated. At the near-side, local
charge conservation in the fragmentation process implies an 
enhanced probability to detect
an associated particle with charge opposite to the
trigger particle. This behaviour has been observed at higher
collision energy~\cite{star2}, in accordance with simulations by 
PYTHIA~\cite{kniege}.

At the away-side, no charge correlations between trigger particle
and associated particle have been
observed at RHIC~\cite{star2}, again in agreement with PYTHIA 
calculations~\cite{kniege}.
We emphasize that this situation is different at SPS energies,
which provides a unique tool to investigate the modification of jet 
properties in A-A via the study of charge correlations, as outlined
in the following.

At large $x_t=2\cdot p_t/\sqrt{s_{\rm NN}}$ particle production
is dominated by valence quark scattering. 
This implies that net
charge conservation becomes relevant in this kinematic range,
leading to correlations among the particles associated to a jet
or di-jet.
Moreover, the large $x_t$ domain at SPS ($\sqrt{s_{\rm NN}}=17.2$~GeV)
is reached at rather moderate $p_t$, like those investigated in
this study, i.e. $p_t\approx3-4$~GeV/$c$.
At such $p_t$, the total number of hadrons produced in the final state
of a $2\rightarrow2$ parton scattering is rather small, 
leading to correlation effects due to {\em global} charge  
conservation in the di-jet system.
A study of simulated "Pb-Au events" at $\sqrt{s_{\rm NN}}=17.2$~GeV
generated with PYTHIA by proper
superposition of pp, np and nn collisions confirms that, at SPS energy,
charge
correlations extend to the away-side. This leads to enhanced correlations
of unlike-sign combinations of trigger and associated particles due
to global charge conservation. Moreover, there is a dominance of 
positive particles
as a consequence of the positive net charge of the valence quarks.
These correlation patterns, characteristic for parton-parton scattering
in elementary collisions at SPS energy, can be used as a reference for
studies in Pb-Au, where modifications or a disappearance of charge
correlations at the away-side may indicate a large degree of dissipation
of the parton energy to the surrounding medium.



\begin{figure}[t]
\centering
\includegraphics[width=0.40\textwidth]{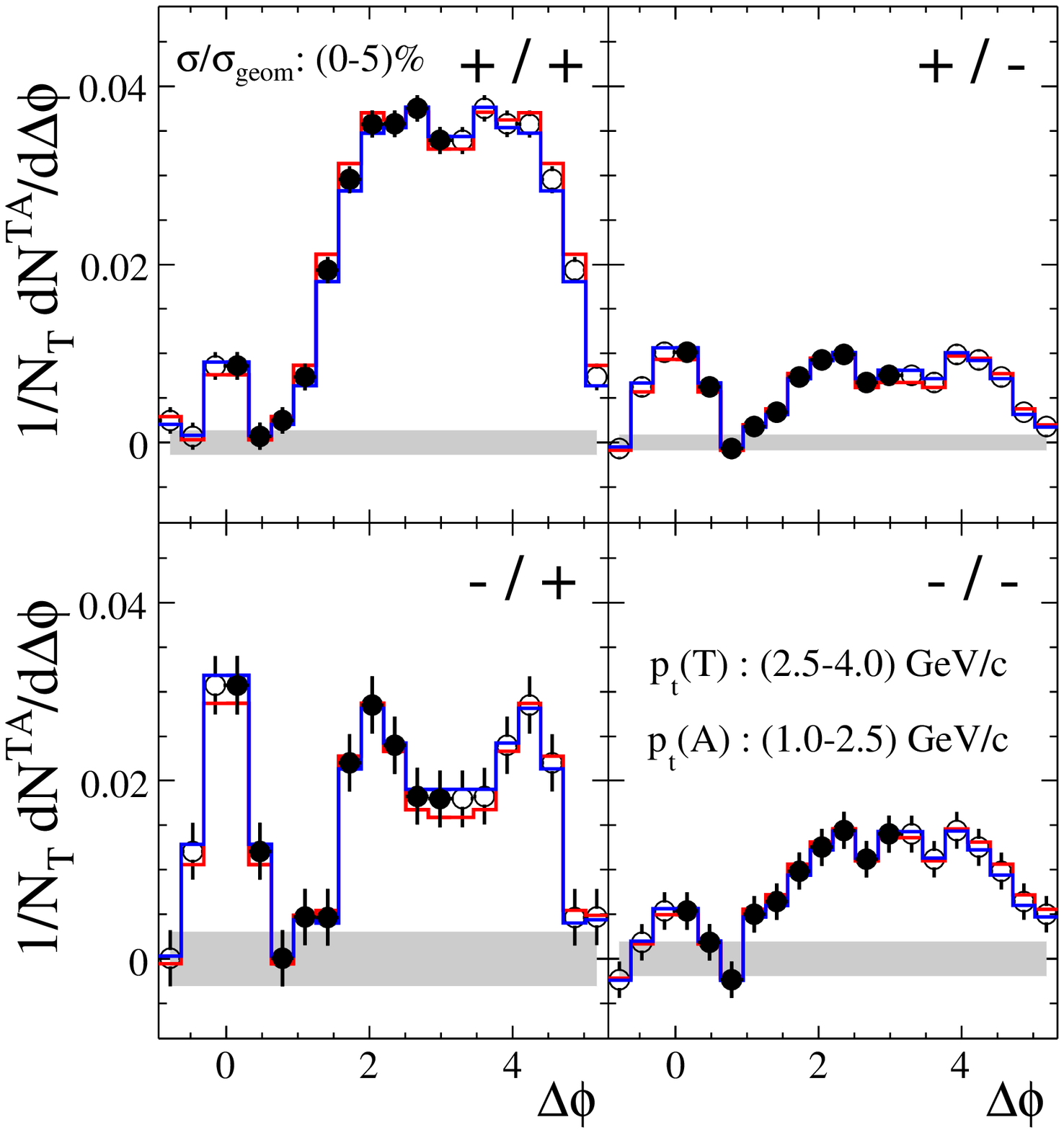}
\includegraphics[width=0.40\textwidth]{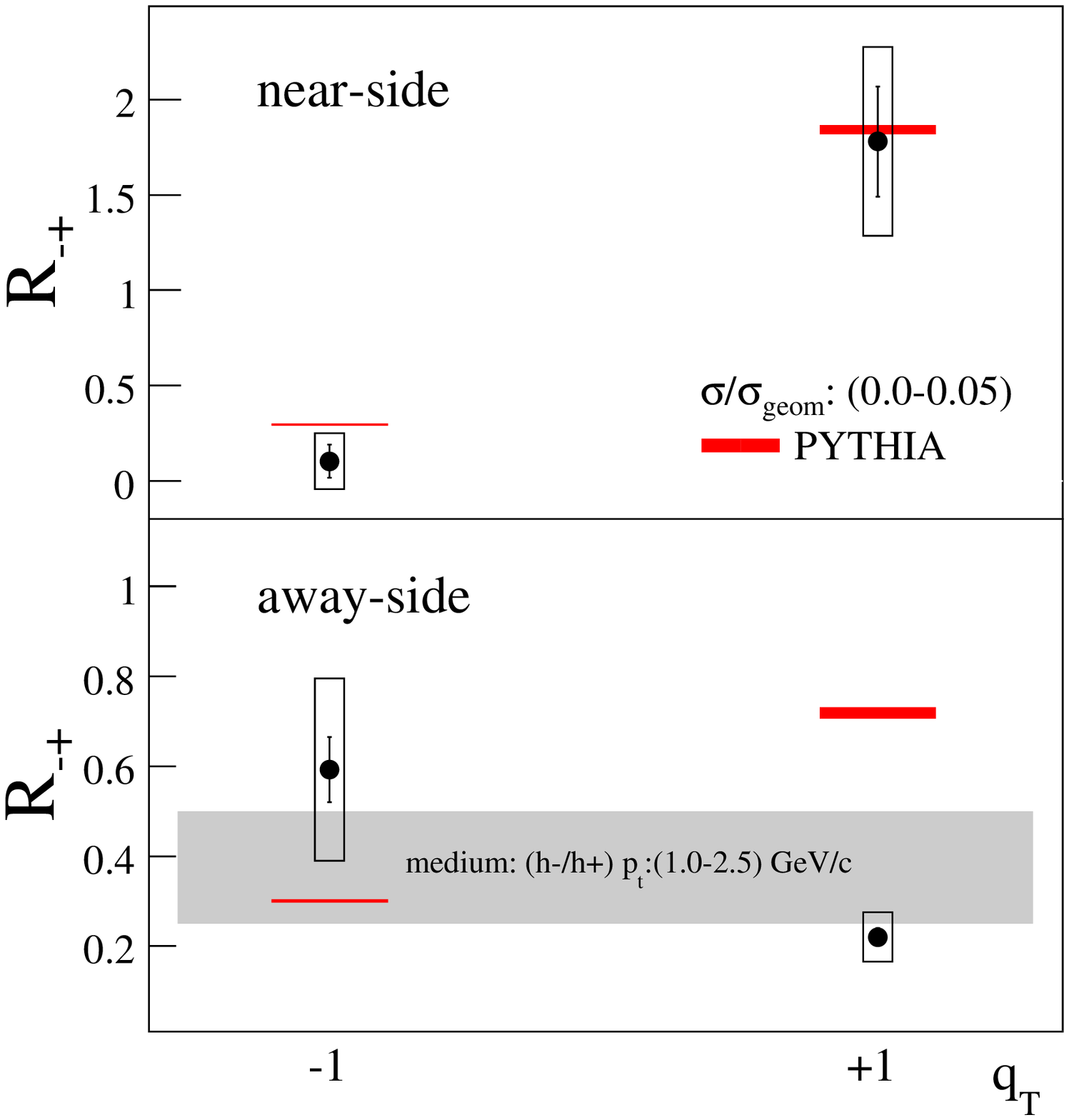}
\caption[]{Left panel: Jet-associated yield in Pb-Au for different 
trigger/associate charge combinations. Right panel: The ratio $R_{-+}$
in comparison to PYTHIA calculations. The inclusive ration $h^-/h^+$ 
is indicated by the grey band (from~\cite{ceres2}).}
\label{fig4}
\end{figure}

In Fig.~\ref{fig4} (left panel) 
the conditional yield for different trigger/associated
particle charge combinations is shown. At the near-side and for a given
trigger charge, we observe a larger yield for the unlike-sign charge 
combinations, consistent with the expectation for jet fragmentation.

In the following we study, for a given trigger charge,
the ratio $R_{-+}$ of the
jet-associated yields of negative over positive associated 
particles~\cite{ceres2}.
The integration limits in $\Delta \phi$ are as stated above. 
We observe very good agreement of $R_{-+}$ at the near-side with
predictions from PYTHIA, see Fig.~\ref{fig4} (right panel), 
corroborating that
jet fragmentation is the origin of the observed correlations. 
In contrast, the data deviate significantly from the PYTHIA 
reference at the away-side, in particular for positive trigger particles.
The data are, however, in good agreement
with the inclusive ratio of negative to positive particles in this
$p_t$ range, indicated by the grey band. This suggests that the 
correlation pattern expected from jet fragmentation in elementary 
collisions is modified as a consequence of significant energy transfer
from the parton to the medium.

In summary, the observed
charge correlation patterns at the near-side give strong evidence that
the origin of high-$p_t$ correlations are jets. At the away-side,
shape and magnitude of the correlation is similar to measurements 
at RHIC. In comparison to PYTHIA calculations, we observe at the
away-side an excess of soft particles and a strong modification
of the charge correlations. Both spectrum and charge composition
of the near-side are consistent with the bulk medium. These observations
are suggestive of significant final state interactions of hard-scattered 
partons with the hot and dense medium formed in heavy-ion collisions at SPS.





\end{document}